\documentclass{iopart}

\usepackage{iopams,setstack}  
\usepackage{graphicx}% Include figure files
\usepackage{bm}% bold math
\usepackage{url}
\usepackage{multirow}

\graphicspath{{figure/}}
\renewcommand{\text}[1]{\mathrm{#1}}

\renewcommand{\vec}[1]{\mathbf{#1}}

\def\be{\begin{equation}}
\def\ee{\end{equation}}
\def\bi{\begin{itemize}}
\def\ei{\end{itemize}}
\def\bea{\begin{eqnarray}}
\def\eea{\end{eqnarray}}
\def\ba{\begin{array}}
\def\ea{\end{array}}

\def\bn{\begin{enumerate}}
\def\en{\end{enumerate}}

\begin{document}

%\baselineskip=2.0\baselineskip

%\preprint{AEI-2007-119}

\title{Sparse sky grid for the coherent detection of gravitational wave bursts} 

\author{Olivier Rabaste}
\ead{Olivier.Rabaste@apc.univ-paris7.fr}
\address{CNRS, AstroParticule et Cosmologie, 10, rue Alice Domon et L\'eonie Duquet, 
75205 Paris Cedex 13, France}
\author{\'Eric Chassande-Mottin}
\ead{Eric.Chassande-Mottin@apc.univ-paris7.fr}
\address{CNRS, AstroParticule et Cosmologie, 10, rue Alice Domon et L\'eonie Duquet, 
75205 Paris Cedex 13, France}
\author{Archana Pai}
\ead{Archana.Pai@aei.mpg.de}
\address{Max-Planck Institut f\"ur Gravitationsphysik, Am M\"uhlenberg 1, 14476 Potsdam, Germany}
\date{\today}

\begin{abstract}
  The gravitational wave detectors currently in operation perform the
  analysis of their scientific data jointly. Concerning the search for
  bursting sources, coherent data analysis methods have been shown to
  be more efficient. In the coherent approach, the data collected by
  the detectors are time-shifted and linearly combined so that the
  signatures received by each detector add up constructively (thus
  improving the resulting signal-to-noise ratio). This operation has
  to be performed over a sky grid (which determines the sky locations
  to be searched). A limitation of those pipelines is their large
  computing cost. One of the available degrees of freedom to reduce
  the cost is the choice of the sky grid. Ideally, the sky sampling
  scheme should adapt the angular resolution associated with the
  considered gravitational wave detector network. As the geometry of
  detector network is not regular (the detectors are not equally
  spaced and oriented), the angular resolution varies largely
  depending on the sky location. We propose here a procedure which
  designs sky grids that permit a complete sky coverage with a minimum
  number of vertices and thus adapt the local resolution.
\end{abstract}
                                
\maketitle

\section{Introduction}
According to Einstein's theory of General Relativity, the
gravitational interaction manifests the geometry of space-time curved
by matter. The theory also predicts the existence of radiative
solutions to the space-time dynamics which are called
\emph{gravitational waves}. Several detectors (including LIGO in the
US, Virgo and GEO in Europe) \cite{detectors} have been built in the
last decade to provide the direct observation of the gravitational
waves. They are mainly long-baseline Michelson-type laser
interferometers able to sense the weak strain in the arms of the
instrument caused by the gravitational waves radiated by astrophysical
sources.

While no gravitational wave has been observed so far by these
detectors, a first detection is plausible with the sensitivities that
will be achieved for the up-coming data takings (labeled S6/VSR2, to
start in 2009). Combining gravitational wave data with other types of
observations may provide crucial help \cite{collaboration08} as they
allow better background rejection. For instance, a follow-up search
for optical counterparts has been suggested in
\cite{kanner08:_looc_up}. The search for coincident high-energy
neutrinos has also been proposed in \cite{Aso:2007wg}.

As these combined searches check the occurence of an (optical or
neutrino) event in time/spatial coincidence with the gravitational
wave triggers, they rely on an estimate for the location of the
gravitational wave source.

Several data analysis pipelines have been proposed to search for bursts
of gravitational waves (typically from star collapses or binary
mergers), our primary focus here. We will be interested in the ones
\cite{Pai:2000zt,klimenko08,Chatterji:2006nh,pai08} which perform the
coherent processing of the data streams collected by each
gravitational wave detectors, as they have shown to be particularly
efficient.

In the coherent approach, the data collected by the detectors are
time-shifted and linearly combined so that the signatures received by
each detector add up constructively if they originate from a given sky
location. The transient signal is then searched for in the combined
data stream.  This operation is repeated by scanning many sky
locations. The result is a \textit{likelihood sky map} which
quantifies the ``likelihood'' of an actual gravitational wave burst
source in a particular sky location. The computation of likelihood sky
maps requires the discretization of the sky sphere. A uniform sky
sampling (see e.g., \cite{Arnaud:2003zq}) is usually used in practice.

A fundamental limitation of the coherent pipelines is their large
computational cost. One of the available degrees of freedom to reduce
the cost is the choice of the sky grid. On one hand, the computing
cost scales linearly with the number of grid points or vertices (the same
calculation is repeated for each bin of the sky grid with different
data and coherent mixing coefficients). On the other hand, the grid
should not be too coarse not to miss any signals. In this paper, we
address this trade-off and propose a \textit{procedure which designs
  a sky grid that permits a complete sky coverage with a minimum number
  of vertices}.

This procedure determines the location of the grid vertices on the
basis of an estimate of the local angular resolution of the
gravitational wave detector network. It thus takes into account both the
specific geometry of the detector network and the characteristics of
the individual detector antenna patterns. Calibration uncertainties
and other timing errors may be also folded in.

We give here a brief outline of the method and connect the various
parts to the corresponding section of the paper. In
section~\ref{s:framework}, we describe the context of this study. We
describe the characteristics of the signature left by a passing
gravitational wave. We spend special attention on the ``travel time''
(difference in the time of arrival of gravitational waves at each
detector with respect to a reference one) and its property because
it is an important ingredient for the sky sampling problem.

The direction of a source can be equivalently described in terms of
the usual spherical coordinates or using the set of travel times. In
section~\ref{s:td_to_sky_grids}, we produce a first sky grid by
sampling regularly in the coordinate system associated to the travel
times. We propose a robust mapping between the discretized
travel times and spherical angles.

The sky grid produced in the first step is ``over-sampled'' as it is
based on timing information only. In a second step presented in
section~\ref{s:min_sample_set}, we use the information of the position 
and orientation of all the detectors and extract the smallest sub-set 
of vertices that ensures complete sky coverage for a given loss in
signal-to-noise ratio. We build this sub-set by
casting the grid size minimization as a \textit{set covering problem}
that can be efficiently solved using a greedy procedure based on a
dual Lagrangian relaxation heuristic. As an illustration, we produce
these sky-grids for an idealized detector network and for the
LIGO-Virgo detector network.  Section~\ref{s:conclusion} concludes
this paper.

\section{Framework}
\label{s:framework}

Let us consider a network of $D$ gravitational wave detectors.  The
response of detector $i$ to an incoming gravitational wave emitted
from the direction $(\phi,\theta)$ can be written as \cite{pai08}:
\begin{equation}
s_i(t)=\Re \left[ F_i(\phi,\theta)^* h(t-\tau_i(\phi,\theta)) \right],
\label{e:received_signal} \hspace{2cm} t \ge t_i
\end{equation}
where $F_i$ is the complex-valued antenna pattern ($x^*$ denotes the
complex conjugate of $x$), and $h(t)=h_+(t)+ih_{\times}(t)$ is the
complex-valued gravitational wave signal combining the two
polarizations $+$ and $\times$. $t_i$ is the time of arrival of
gravitational wave at detector $i$. In eq. (\ref{e:received_signal}),
time is defined with respect to a reference which we arbitrarily set to
be the time at detector $1$.  With this convention,
$\tau_i(\phi,\theta) = (t_i - t_1)$ denotes the time taken by the
wave to travel from detector $1$ to detector $i$.
For this reason, we refer to
$\tau_i(\phi,\theta)$ as \textit{travel time}. It is proportional
to the projected distance between detectors $1$ and $i$ onto the
projection direction of the wave, viz.
\begin{equation}
\label{eq:taui}
\tau_i(\phi,\theta)=\frac{1}{c} (\vec{r}_i-\vec{r}_1)^T \vec{w}(\phi,\theta),
\end{equation}
where $\vec{r}_i$ is the coordinate vector of detector $i$,
$\vec{w}(\phi,\theta)$ is the unit wave vector and $c$ is the speed of
light in free space. Here, bold symbols are (column) vectors and
$\vec{x}^T$ denotes the transpose of $\vec{x}$.

\subsection{Link between spherical coordinates and travel times}
\label{sec:sc_to_tt}

\begin{figure}
\begin{center}
 \includegraphics[width=.42\textwidth]{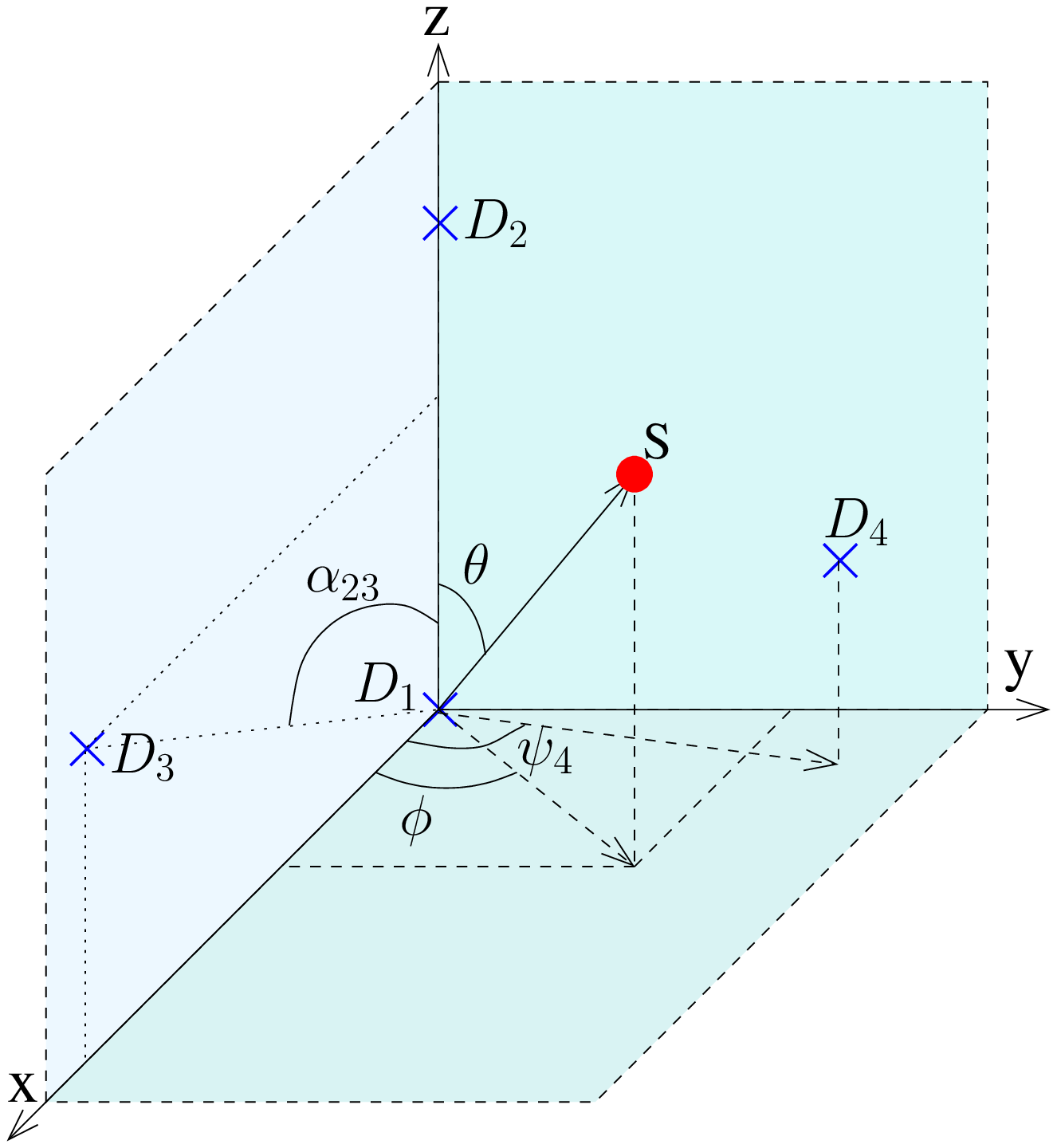}
\end{center}
\caption{\textbf{Reference frame} --- We choose here to work in a
  reference frame (described for example in \cite{bhawal95}) in which
  expressions linking spherical coordinates and travel times are
  simple. In this graph, $D_i$ corresponds to the position of detector
  $i$. The origin of the network is chosen to be detector $D_1$ which
  also define our time reference. The $z$ axis is defined as the line
  joining $D_1$ and $D_2$. The $x$ axis is defined to be the
  intersection between the plane orthogonal to the $z$ axis and the
  plane containing $D_1$, $D_2$ and $D_3$. Finally the $y$ axis is
  chosen so that the $x$, $y$ and $z$ axes form a right-handed
  coordinate system. $D_1S$ represents the direction of the
  gravitational wave source.}
\label{fig:network_frame} 
\end{figure}

In this section, we describe the relationship between the spherical
coordinates $(\phi,\theta)$ and the set of travel times
$\{\tau_i(\phi,\theta) \}_{i=1,\dots,D}$. This has already been
examined in \cite{bhawal95} and this section summarizes
some of the results presented in this article.

We choose to work in the reference frame presented in
Fig.~\ref{fig:network_frame} where the expressions linking spherical
coordinates and travel times are simple.  Let $T_i=d_{1i}/c$ be the
time of flight between detectors $D_1$ and $D_i$ (here
$d_{1i}=\|\vec{r}_i-\vec{r}_1\|$ is the distance between $D_1$ and
$D_i$).  With these notations, the travel time can be expressed with
respect to the spherical coordinates $(\phi,\theta)$ in the reference
frame by:
\begin{equation}
\label{eq:traveltime}
\tau_i = -T_i \left( \cos \theta \cos \alpha_{2i} + \sin \alpha_{2i} \cos \left( \phi-\psi_i \right) \sin \theta  \right),
\end{equation}
where $\alpha_{2i}$ is the angle between the lines $(D_1D_2)$ and
$(D_1D_i)$ and $\psi_i$ is the angle between the $x$ axis and the line
$(D_1D_{ixy})$ (with $D_{ixy}$ being the projection of $D_i$ onto the $xy$
plane).

The spherical coordinates in this frame can be obtained by the
reciprocal expressions:
\begin{eqnarray}
\theta &= \cos ^{-1} \left( -\frac{\tau_2}{T_2}\right),\label{eq:theta}\\
\phi &= \pm \cos ^{-1} \left( -\frac{T_2 \tau_i - T_i \tau_2 \cos \alpha_{2i}}{T_i\sqrt{T_2^2-\tau_2^2} \sin \alpha_{2i}}\right),
\label{eq:phi}
\end{eqnarray}
with $i \geq 3$. These equations display well-known features of
triangulation: the zenith angle $\theta$ can be determined
only by two detectors, however the azimuth angle $\phi$ requires at
least four detectors to be completely determined (including the sign).
Thus alternatively, the set of travel times may be seen as ``coordinates''
equivalent to the spherical angles.

As we will describe later in Sec. \ref{s:td_to_sky_grids}, the
coordinate system generated by the collection of travel times is more
convenient to produce sky grids adapted to the considered detector
network. However, this coordinate system is not ``standard''. This is
why we examine its geometry into some more details now. The 2-sphere
is completely characterized by the two spherical angles. For a 
number $D$ of detectors, the number of travel times is $D-1$.
We need to identify the corresponding 2-sphere in the resulting $D-1$-
dimensional coordinate space.

From eq. (\ref{eq:traveltime}), we see that the travel time $\tau_i$
takes values in $[-T_i,T_i]$. Therefore, the admissible values for the
coordinates $\pmb{\tau}=[\tau_2, \ldots, \tau_D]^T$ are located 
inside a $D-1$-rectangular cuboid. But, for a detector network with $D>2$, the
travel time coordinates do not span this cuboid entirely. For instance,
it is impossible to have $\tau_2=T_2$ and $\tau_3=T_3$ unless $D_1$,
$D_2$ and $D_3$ lie on a straight line. From the constraints imposed
by the physics, i.e. the propagation of the wave and the geometry of
the network, it is possible to determine the \textit{admissible
  surface} spanned by the travel time coordinates.

The equation of this two dimensional admissible surface can be
obtained from 
Eq.~(\ref{eq:traveltime}) \cite{bhawal95}. It can be written in the 
form 
\begin{eqnarray}
\label{eq:adm_surf}
\pmb{\tau}^T \vec{A}_D \pmb{\tau} &\le& B_D  \hspace{2cm} D=3 \, \\ 
&=& B_D \hspace{2cm} D=4 \,. 
\end{eqnarray}
The expressions
of the (symmetric) matrix $\vec{A}_D$ and $B_D$ were obtained in
\cite{bhawal95} for $D=3$ and $D=4$. We give them below for
completeness.

For $D=3$, the admissible surface is an \textit{ellipse} with
\begin{eqnarray}
\vec{A}_3&=
\left[
 \begin{array}{cc}
 \frac{T_3^2}{T_2^2} & -\frac{T_3}{T_2} \cos \alpha_{23} \\
 -\frac{T_3}{T_2} \cos \alpha_{23} & 1
 \end{array}
\right],\\
B_3&=T_3^2 \sin^2 \alpha_{23}.
\end{eqnarray}

For $D=4$, the admissible surface is an \textit{ellipsoid}. 
If $\vec{A}_4 \equiv (A)_{ij}$, we then have
\begin{eqnarray*}
A_{11} &=&  \frac{T_4^2}{T_2^2} \left[ \left( \cos \alpha_{24} - \sin \alpha_{24} \cos \psi_4 \cot \alpha_{23} \right)^2 + \left(\frac{\sin \alpha_{24}}{\sin \alpha_{23}} \sin \psi_4\right)^2 \right], \\
A_{22} &=&  \left( \frac{T_4 \sin \alpha_{24}}{T_3 \sin \alpha_{23}} \right)^2, \\
A_{33} &=& 1,\\
A_{12} &=&  \frac{T_4^2}{T_2 T_3} \frac{\sin \alpha_{24}}{\sin \alpha_{23}} \left( \cos \alpha_{24} \cos \psi_4 - \sin \alpha_{24} \cot \alpha_{23} \right), \\
A_{13} &=&  - \frac{T_4}{T_2} \left( \cos \alpha_{24} - \sin \alpha_{24} \cos \psi_4 \cot \alpha_{23} \right),\\
A_{23} &=&  -\frac{T_4}{T_3}\frac{\sin \alpha_{24}}{\sin \alpha_{23}} \cos \psi_4,  
\end{eqnarray*}
while $B_4=T_4^2\sin^2\alpha_{24} \sin^2\psi_4$.

With the projection of the ellipsoid onto the ($\tau_i,\tau_j$)
plane, we retrieve the ellipse for the corresponding $D_1$, $D_i$ and $D_j$
three-detector network. The two points from the two shelves of the
ellipsoid which get projected in a single one are associated to the
two possible values for $\phi$ in Eq.~(\ref{eq:phi}).

For $D>4$ detectors, it is not straightforward to determine the
equation of the admissible surface. However, we notice that any
triplet taken from the set $\{ \tau_i(\phi,\theta) \}_{i=2,\dots,D}$
must verify the corresponding ellipsoid equation. The admissible
surface is thus the intersection of $C^{D-1}_3$ ellipsoidic cylinders.
Those equations are redundant. As a sky location is associated to one
travel time triplet only, all travel times can be determined from
three of them. We need one equation to get one ``master'' travel time
triplet and $D-1-3$ equations to determine the remaining travel times,
thus a total of $D-3$ equations.

\section{Sampling the travel time space and  mapping to sky grid}
\label{s:td_to_sky_grids}

The coherent analysis follows directly from the computation of the
``global'' or ``network'' likelihood ratio testing the presence of a
signal in the output from all detectors jointly. The network
likelihood ratio involves linear combination of the data streams such
that the response of the detectors to an incoming gravitational wave
adds coherently.  This implies the adjustment of the data streams in
time and phase \cite{pai08,klimenko08,Chatterji:2006nh}. \textit{Time
  delays} are applied to ``synchronize'' the various responses,
i.e. compensate the travel time of the wave between
detectors. Clearly, these time delays are intimately related to the
travel time coordinates introduced in the previous section.

Following the principle of the generalized likelihood ratio test, in
presence of unknown parameters, the network likelihood ratio (more
precisely its $-\log$) is maximized over those parameters. Here, the
parameters are the coordinates of the source location $\phi$ and
$\theta$ and the parameters connected to the physical
characteristics of the source (e.g., orientation of the orbital plane
for an inspiralling binary, masses of the binary stars)
and to the waveform morphology (e.g., central
frequency for a sine-Gaussian type bursts).

We are interested here in the maximization over $\phi$ and $\theta$.
Mathematically, this is a non-linear optimization problem (due to the
adjustment of the time delays in particular). The standard approach
for its resolution is to find the maximum of the network likelihood
ratio over a sky grid. Instead of the usual scheme based on the
uniform discretization of the spherical coordinates, we propose here
to discretize (the admissible surface of) the travel time coordinates
introduced in the previous section. The resulting sky grid is built
using (at least part of) the geometry of the detector network (i.e.,
the relative position of the detectors) and is thus more adapted to
the problem at hand. The procedure goes in two steps. We first present
the adequate sampling of the travel times. Then we map the sampled
travel times to actual sky locations.

\subsection{Sampling the travel time admissible surface}
\label{s:sampling_admissible_surface}

The data streams at the detectors are sampled at a given sampling
frequency $f_s$. We propose here to \textit{sample the travel time
  coordinates with the same pace}. This presents the great advantage that
the time delays applied in the coherent analysis are an integer number
of samples. No interpolation procedure is thus needed to compute the
coherently combined data streams.

In the following, we will label with a superscript $s$ the sampled
variables. Let $t_i^s$ denote the closest time sample from time $t_i$
at detector $i$. We have $t_i^s=t_if_s+\epsilon_i$ with $\epsilon_i
\in (-0.5,0.5)$ the truncation error.

The travel time being the time difference $\tau_i=t_i-t_1$, its
sampled version $\tau_i^s=t_i^s-t_1^s$ is affected by two truncation
errors since $\tau_i^s=\tau_i f_s+\epsilon_i-\epsilon_1$.
Now considering the entire set of coordinates defined in
Sec. \ref{sec:sc_to_tt}, we have
\begin{equation}
\label{eq:sampledtau}
\pmb{\tau}^s=\pmb{\tau} f_s+\pmb{\epsilon}-\epsilon_1 \vec{1}
\end{equation}
where $\vec{1}$ is a vector with all the entries equal to $1$.

For all $\pmb{\tau}$ on the admissible travel time surface, we want to
find the integer vectors $\pmb{\tau}^s$ provided that $\epsilon_1$ and
the components of $\pmb{\epsilon}$ are in $(-0.5,0.5)$. Let us first
assume that $\epsilon_1=0$. We accept $\pmb{\tau}^s$ if the admissible
surface intersects the cube of edge $1$ centered on $\pmb{\tau}^s$.
Now, consider that $\epsilon_1\neq 0$.  We accept $\pmb{\tau}^s$ if
there exists $\epsilon_1 \in (-0.5,0.5)$ such that the admissible
surface translated by $\epsilon_1 \vec{1}$ intersects the cube of edge
$1$ centered on $\pmb{\tau}^s$. This method ensures that the boundaries 
of the admissible surface get
well sampled. This precaution is essential for four and more detector
case since \textit{all points} of the admissible surface belong to 
the boundary.

\begin{figure}
\begin{center}
  \includegraphics[width=.45\textwidth]{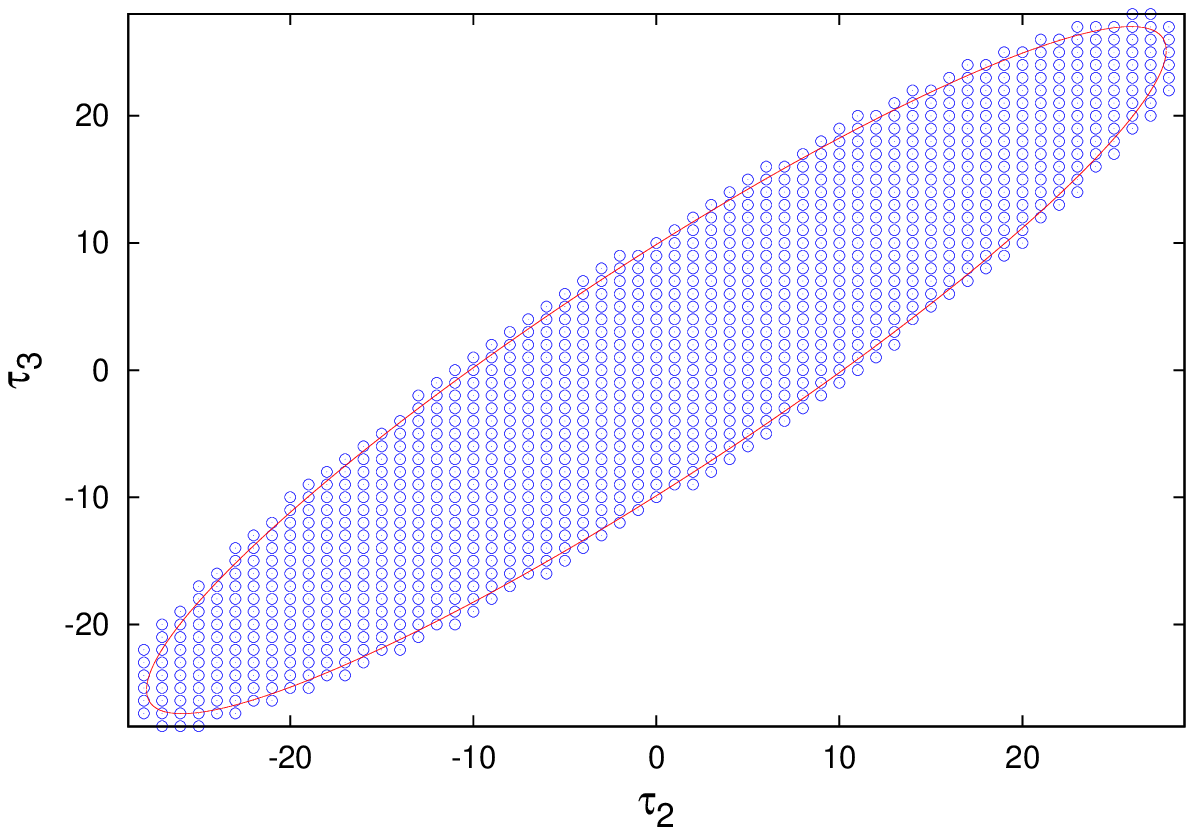}
  \includegraphics[width=.45\textwidth]{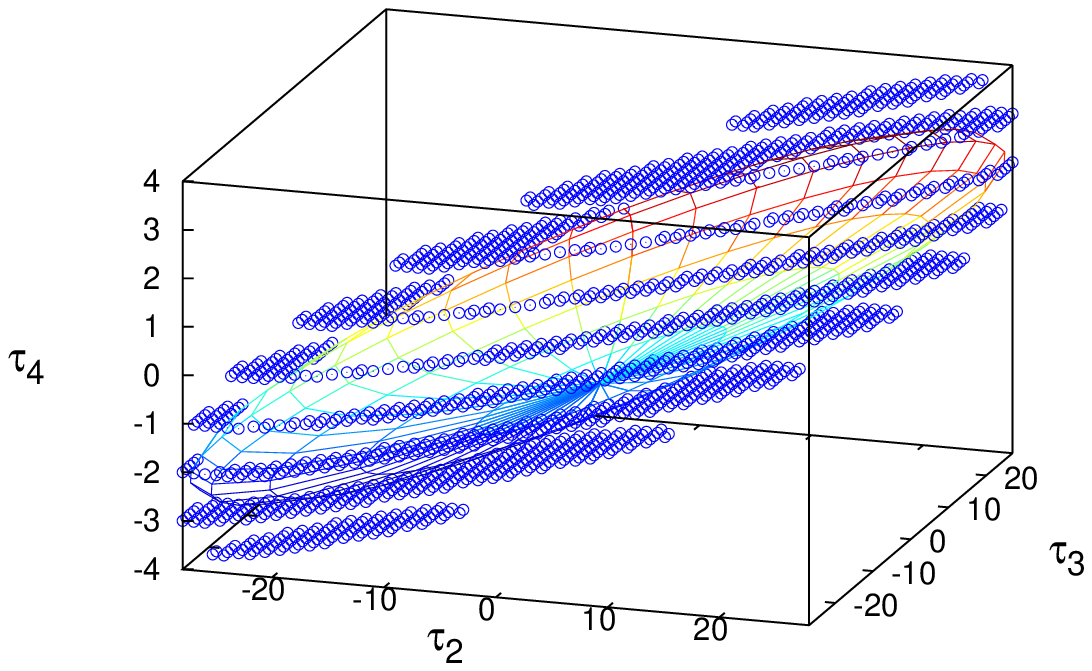}
\end{center}
\caption{\textbf{Admissible surfaces of travel times} for a three
  (left, ellipse) and four (right, ellipsoid) detector network and
  their sampling ($f_s=$ 1024 Hz). The
  unit is milliseconds. The positions of the detectors corresponds to
  the actual positions of Virgo (Pisa, Italy), LIGO H (Hanford, WA,
  US) and LIGO L (Livingstone, LA, US) with GEO (Hannover, Germany)
  added for the four detector case. The sample points obtained with
  the method described in Sec. \ref{s:sampling_admissible_surface} are
  indicated with blue dots. Part of the points appear to fall slightly
  off the admissible surface due to round-off (see discussion in
  Sec. \ref{sec:td_sphangles}).}
\label{fig:samples_fs_1024_VHL} 
\end{figure}

We demonstrate this method for $D=3,4$ detector network as illustrated
in Figure~\ref{fig:samples_fs_1024_VHL}. For a three detector network,
the admissible surface is an ellipse. Most of the admissible samples
lie inside the ellipse for which the proposed method is very
efficient.  For D=4, the admissible surface is an ellipsoid which
makes the admissibility criterion difficult to verify. This is mainly
because all of the admissible points lie on the boundary and the
criterion amounts to determining the intersection of an ellipsoid and
a 3-cube. However we simplify the selection criterion by looking for
cubes that have at least one vertex inside the ellipsoid and at least
one outside\footnote{It is possible to refine this criterion by
  considering more points on the surface of the cube instead of the
  vertices only}. The result of this simplified method is presented in
Figure~\ref{fig:samples_fs_1024_VHL}.

For larger detector network, this basic method has to be abandoned since
the admissible surface is not known analytically. One possibility is
to use Monte Carlo (MC) trials: generate a random set of uniformly
distributed sources, compute the wave arrival times and collect the
list of distinct discretized travel times. The MC method has the
additional advantage that the samples with a very low probability of
appearance, i.e. samples with a small intersection between the sample
cube and the set of translated admissible surface, are automatically
dismissed.

Table~\ref{t:number_samples} gives a comparison of the different
discretization methods. As expected, taking the physical properties of
travel times (the admissible surface) into account decreases
significantly the number of samples. Also the larger the detector
network, the more
samples are discarded by the MC method. This means that the number of
very unlikely samples increases with the number of detectors. Note
finally that the proportion of samples being dropped by the MC method
increases with the sampling frequency.

% figure below obtained with check_get_admissible_index_3det.m
% figure below obtained with check_get_admissible_index_4det.m

% data below obtained with check_get_admissible_index_3det.m, check_admissible_index_4det.m for the exhaustive values and run_get_most_probable_time_delays.sh for the most probable ones
\begin{table}
\centering
\begin{tabular}{|c|c|c|c|c|c|}
\hline
\multicolumn{2}{|c|}{$f_s$ (Hz)} & 512 & 1024 & 2048 & 4096 \\
\hline
\hline
\multirow{3}{*}{A: 3-det} & $N_{c}$ & 841 & 3,249 & 12,543 & 49,275 \\
\cline{2-6}
& $N_{a}$ & 285 & 1,002 & 3,726 & 14,341 \\
\cline{2-6}
& $N_{MC}$ & 285 & 1,001 & 3,723 & 14,336 \\
\hline
\hline
\multirow{3}{*}{B: 4-det} & $N_{c}$ & 4,205 & 29,241 & 188,145 & $1.43 \times 10^6$  \\
\cline{2-6}
& $N_{a}$ & 869 & 4,101 & 22,611 & 143,155 \\
\cline{2-6}
& $N_{MC}$ & 866 & 3,983 & 16,650 & 66,454 \\
\hline
\hline
\multirow{2}{*}{C: 5-det} & $N_{c}$ & 966,735 & $1.35 \times 10^7$ & $1.96 \times 10^8$ & $3.07 \times 10^9$\\
\cline{2-6}
& $N_{MC}$ & 5,725 & 22,985 & 90,740 & 351,171 \\
\hline
\end{tabular}
\caption{\textbf{Comparison of various for the discretization schemes used for the travel time space.} ---
  $N_c$, $N_a$ and $N_{MC}$ are the total number of grid points of the travel time grids obtained (i) if their physical properties
  are ignored (i.e., uniform sampling of $\tau_i$ between $[-T_i,T_i]$), (ii) if we sample the 
  admissible surface with the method described in Sec.~\ref{s:sampling_admissible_surface}, 
  (iii) from the Monte-Carlo search  described in Sec.~\ref{s:sampling_admissible_surface}, resp. 
  The network A is composed of Virgo, LIGO H and LIGO L.
  The network B is composed of these 3 detectors plus GEO (Hannover, Germany).
  The network C is composed of  these 4 detectors plus a detector located in Japan 
  (assumed to be at the location of the decommissioned detector TAMA.)}
\label{t:number_samples}
\end{table}

\subsection{From travel time coordinates to spherical angles}
\label{sec:td_sphangles}

If physically admissible, travel time coordinates can be mapped into
spherical coordinates using Eqs. (\ref{eq:theta}) and (\ref{eq:phi}).
Some of the travel time samples fall close but outside the admissible
surface because of the sampling round-off error (see
Fig.~\ref{fig:samples_fs_1024_VHL}).  While this seems
marginal in the 3-detector case (because it happens to the small
number of points at the boundary of the admissible ellipse), it is the
case for almost all samples in the 4-detector case. A procedure is
needed to map points that are slightly off the admissible surface to a
sky position. We investigate this question here.

Assuming a given $\pmb{\tau}^s$, we search for the closest
$\pmb{\tau}$ belonging to the travel time admissible surface $\mathcal{S}$:
\begin{equation}
\pmb{\tau}^{\star}=\arg \min_{\pmb{\tau} \in \mathcal{S}} \|
\pmb{\tau}^s f_s - \pmb{\tau} \|^2_{\vec{X}} \equiv
\mathcal{P}_{\mathcal{S}}(\pmb{\tau}^s f_s),
\label{eq:tau_star}
\end{equation}
where $\|x\|_{\vec{X}}$ is the norm associated with the inner product
induced by the positive definite matrix
$\vec{X}$. $\mathcal{P}_{\mathcal{S}}(\cdot)$ is thus the projector
onto $\mathcal{S}$ defined by this inner product.

We may opt for a standard least-square minimization by setting
$\vec{X}$ to identity. In a more accurate version, we may also model
the timing errors $\epsilon_i$ in Eq. (\ref{eq:sampledtau}) as
Gaussian random variables with zero mean and variance $\sigma_i^2$ and
use $\vec{X} = \mathrm{diag} [\sigma_2^2,\ldots, \sigma_D^2] +
\sigma^2_1 \vec{1}\vec{1}^T$. In this last case, $\pmb{\tau}^{\star}$ in Eq. (\ref{eq:tau_star}) is
the maximum likelihood estimator of $\pmb{\tau}$ (and associate sky
location). The use of this estimator does not restrict to the present
sky sampling problem, but it is also useful in the standard
triangulation scheme \cite{markowitz_zanolin} used to obtain the
source position from a set of measured arrival times. Note that the
model (i.e., the values of $\sigma_i$) may be tuned to integrate
various timing errors (such as calibration uncertainties).

Let us consider networks with $D=3$ or $4$ detectors.  In the case
where $\vec{X}=\vec{I}$ and including the quadratic expression of
$\mathcal{S}$ in Eq. (\ref{eq:adm_surf}), we are led to the following
constrained minimization problem
\begin{equation}
\label{eq:cstr_min}
\min_{\pmb{\tau}} \pmb{\tau}^T\pmb{\tau}-2(\pmb{\tau}^s f_s)^T\pmb{\tau}
 \mathrm{~~subject~to~~} \pmb{\tau}^T\vec{\mathcal{A}}\pmb{\tau} = 1
\label{e:projection}
\end{equation}
where $\vec{\mathcal{A}}=\vec{A}_D/B_D$. The case with an arbitrary
$\vec{X}$ can be re-casted into this one by setting
$\pmb{\tau}'=\vec{X}^{1/2}\pmb{\tau}$ and
$\vec{\mathcal{A}}'=(\vec{X}^{-1/2})^T \vec{\mathcal{A}}
\vec{X}^{1/2}$. This problem can be solved with the method of Lagrange
multipliers. It follows that the solution is
$\pmb{\tau}=(\vec{I}+\lambda \vec{\mathcal{A}})^{-1}\pmb{\tau}^sf_s$
where the Lagrange multiplier $\lambda$ satisfies the polynomial
equation:
\begin{equation}
\label{eq:lagrange}
(\pmb{\tau}^sf_s)^T(\vec{I}+\lambda \vec{\mathcal{A}})^{-T}\vec{\mathcal{A}}(\vec{I}+\lambda \vec{\mathcal{A}})^{-1}\pmb{\tau}^sf_s=1.
\end{equation}

This equation can be solved by standard root finding algorithms (in
the 3-detector case, Eq.~(\ref{eq:lagrange}) is a fourth-order
polynomial that can be solved analytically by radicals with Ferrari's
method).

This methodology has to be adapted for networks with $D\geq 5$
detectors. As discussed in Sec. \ref{sec:sc_to_tt}, the admissible
surface is described by a set of (at most) $D-3$ quadratic equations
(each corresponding to the ellipsoid equation for a travel time
subset).  Consequently, the number of constraints imposed in the
minimization problem of Eq.~(\ref{eq:cstr_min}) is increased to $D-3$.
This problem cannot be solved using the Lagrangian multipliers as
before. It can however be viewed as a quadratic programming problem
with non-convex quadratic constraints for which numerical methods have
been developed \cite{anstreicher00,lasserre01,henrion03}.

\begin{figure}
\begin{center}
  \includegraphics[width=.45\textwidth]{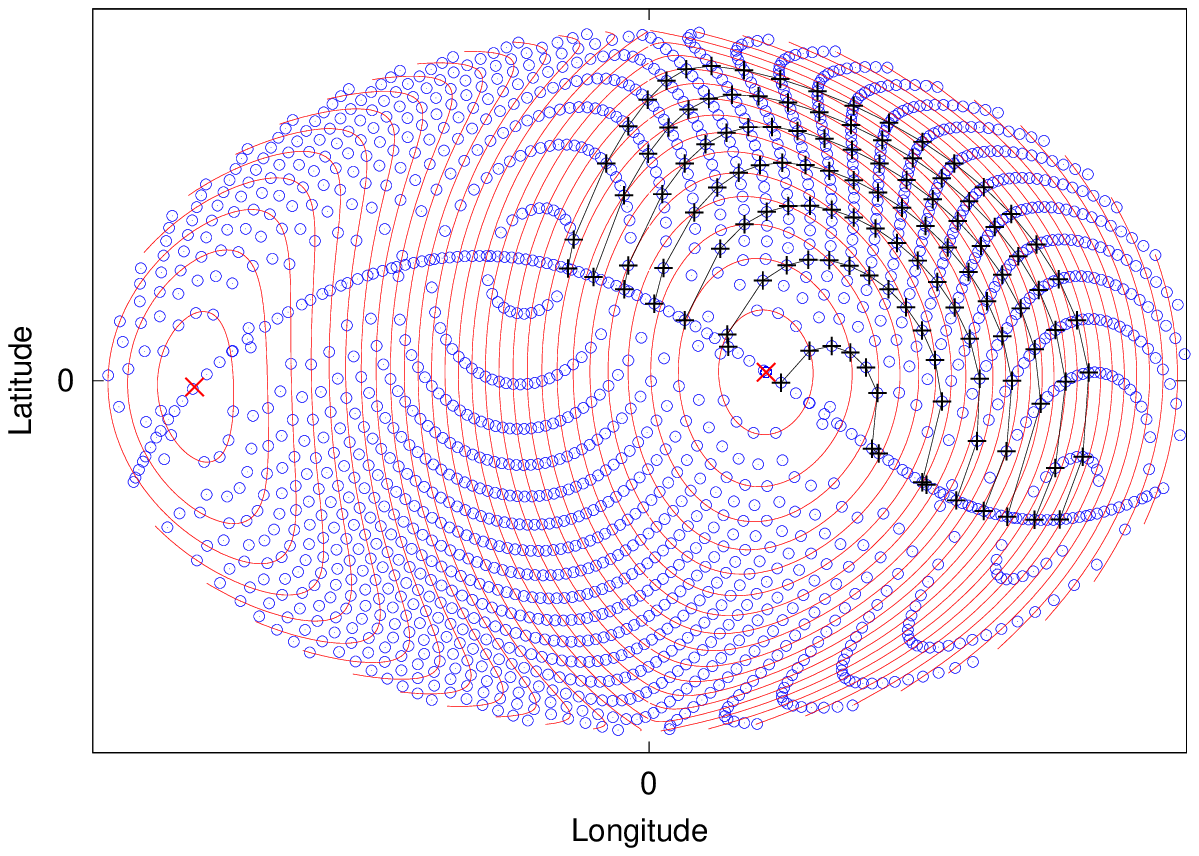}
  \includegraphics[width=.45\textwidth]{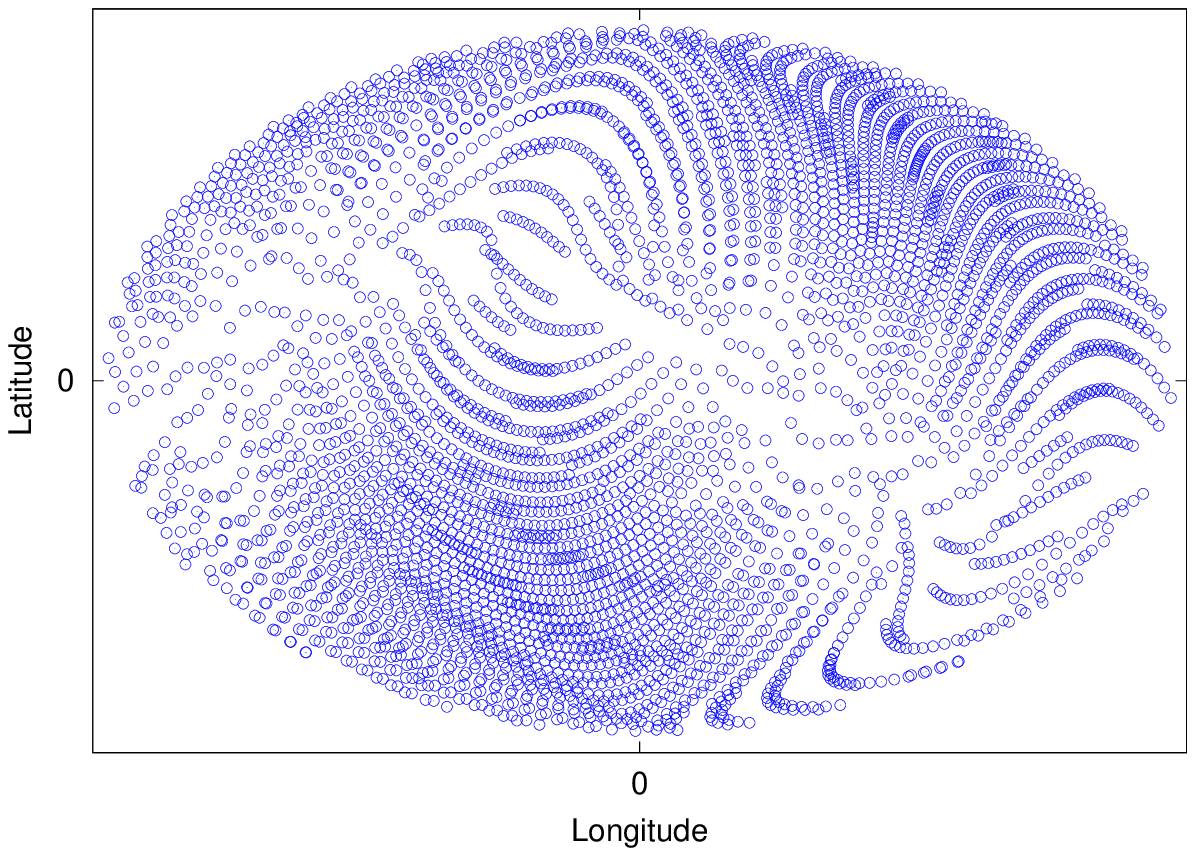}
\end{center}
\caption{\textbf{Sky grids associated to the sampling of the
    admissible travel time surface} presented in
  Fig.~\ref{fig:samples_fs_1024_VHL} obtained by the robust mapping
  procedures described in Sec. \ref{sec:td_sphangles}. The grid
  obtained for the 3-detector network (left) shows a much more
  structured distribution than the one obtained for the 4-detector
  network (right). This structure can be briefly explained as follows:
  first, the ``sinusoidal'' line which splits the figure in two halves
  corresponds to the detector plane ( $\phi=0$ in the network frame,
  see Fig.~\ref{fig:network_frame}). The two crosses $\times$ (in red)
  indicate the points $(\phi,\theta)=(0,0)$ and $(0,\pi)$.  The grid
  results from the intertwinement of several collections of concentric
  circles: the first set of circles is contours of the polar angle
  $\theta$ (or equivalently $\tau_2$, indicated in red), the second
  refers to contours of $\tau_3$ (semicircles shown with black $+$)
  and the third set is formed by contours of
  $\tau_2=\tau_3+C^{st}$. The two sky points along which the third set
  is centered correspond to the minimum and maximum values of
  $C^{st}$.  Geometrically, for a special case of VHL where $T_2 \sim
  T_3$, these 2 sky points correspond to the sky directions along the
  axis (in the LHV plane) perpendicular to the angle bisector
  $D_2-D_1-D_3$ (or L-V-H). It mimics the geometry of the sphere with
  these two points acting as poles and the contour $\tau_2=\tau_3$
  acts like an equator. In any three detector case, this equator is
  defined by $\phi=0$ and $\theta=\tan^{-1} (T_2-T_3
  \cos\alpha_{23})/(T_3 \sin \alpha_{23})$.}
\label{fig:sample_dist_ellipse_fs_1024_constant_theta_circles} 
\end{figure}

\begin{figure}
\begin{center}
  \includegraphics[width=.5\textwidth]{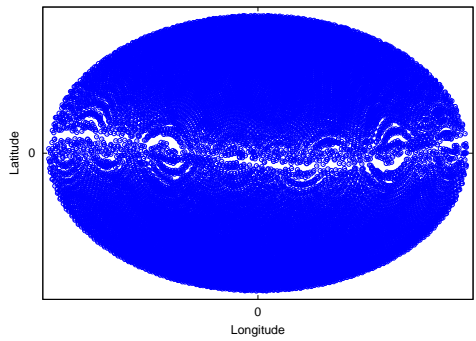}
\end{center}
\caption{\textbf{Sky grid associated to the sampling of the admissible surface} for
  the five-detector network C (see Table \ref{t:number_samples}) and
  assuming $f_s=1024$ Hz. Clearly, the sky sampling is much tighter
  than the one obtained for other networks. This is due to the far
  location (Japan) of the additional detector. The information given
  by the time delays between this detector and the others is very
  accurate and thus valuable.}
\label{fig:sample_dist_5_det_fs_1024} 
\end{figure}

We have applied these mapping algorithms to the travel time grids
presented in Fig. \ref{fig:samples_fs_1024_VHL} with results in
Fig.~\ref{fig:sample_dist_ellipse_fs_1024_constant_theta_circles} and
to a five-detector network with result in
Fig.~\ref{fig:sample_dist_5_det_fs_1024}. Note that the spherical
maps in these figures as well as all the other presented in this
article are obtained using a Hammer-Aitoff projection and using as
reference frame the geographic coordinate system.

As expected the sampling is not \textit{uniform} as it depends on the
network geometry which is \textit{irregular}.  In particular, the grid
gets loose in the vicinity of detector plane ($\phi=0$ in the
reference frame in Fig.~\ref{fig:network_frame}).  The 3-detector sky
grid \footnote{For the 3-detector network, the grid points that are
  strictly inside the admissible ellipse have been mapped to the two
  opposite sky directions owing to the sign ambiguity in
  Eq.~(\ref{eq:phi}), whereas the points located on and outside the
  boundary of the ellipse (they correspond to $\phi=0$ i.e., no sign
  ambiguity) are mapped to only one sky direction.} appears to be
much more structured than the 4-detector one. An explanation of these
structures is given in the caption of
Fig.~\ref{fig:sample_dist_ellipse_fs_1024_constant_theta_circles}.

% figure below obtained with plot_sky_samples_with_concentric_circles.m and using ExhaustiveFile==1 in it  

\section{Smallest grid with complete sky coverage}
\label{s:min_sample_set}

In the previous sections, we have shown how to produce a sky grid from
the discretization of the travel time/time delay space. As this grid
is built from timing information only, it does not use all the
available information (i.e., detector orientation and aperture). The
consequence is that this grid is ``oversampled''. Its large size
generally prevents coherent searches to be performed in real time with
current computers. We want to extract from the sky grid the smallest
subset ensuring the entire sky coverage. This means that we want that
the loss due to the mismatch between any source direction and the
closest grid point be not too large. In this section, we describe 
the selection procedure of the sky grid which tightly
covers the sky sphere with the minimal number of grid points.
Our procedure results in selecting the grid points in areas of the 
sky where the position estimation is accurate, and withdrawing samples 
in areas where the accuracy is poor. 

\subsection{Angular resolution from the network statistic expansion}
\label{s:angular_resol}

Coherent analysis follows directly from the computation of the network
likelihood ratio testing the presence of a signal in the output from
all detectors jointly. The likelihood ratio is a function $L$ of the
data and of the unknown parameters that characterise the expected
signal. According to the principle of the generalized likelihood ratio
test \cite{poor88}, the network likelihood ratio (its $-\log$,
precisely) is maximized over the parameters. We assume here that this
maximization has been done for all parameters but the source
coordinates $\phi$ and $\theta$, which leads to the likelihood sky map
$L(\phi,\theta)$. When a signal is present, it is expected that
this function peaks where the source lies. In this section, we
investigate the width of the peak.

Let us consider that there is a GW source in the direction
$(\phi_0,\theta_0)$.  We obtain the peak width by the Taylor expansion
of $L(\phi,\theta)$ about $(\phi_0,\theta_0)$:
\[
L(\phi,\theta) \approx L(\phi_0,\theta_0) + 
\frac{1}{2} \sum_{\alpha,\beta}\partial^2_{\alpha\beta}L|_{\phi_0,\theta_0} \Delta \alpha \Delta \beta
\]
where $\alpha,\beta$ = $\phi$ or $\theta$ in the summation and we have
$\Delta \phi = \phi-\phi_0$ and $\Delta \theta = \theta-\theta_0$.

In the case where the incident wave\footnote{We assume here that the
  wave is circularly polarized. However, it is straightforward to
  accommodate cases where polarization is elliptical e.g., inspiralling
  binary with arbitrary inclination w.r.t. the line of sight.}
$h(t)=A(t)\exp i \varphi(t)$ is a quasi-periodic burst of amplitude
$A(t)$ and phase $\varphi(t)$, the likelihood ratio is expressed
as \cite{Pai:2000zt}
\begin{equation}
\label{eq:LR}
L(\phi,\theta) = \sum_{n,m=1}^D P_{nm} C_n C^*_m,
\end{equation}
where the correlation measurement\footnote{For simplicity, we give the
  expression of $C_n$ assuming white Gaussian noise. It is
  straightforward to extend this expression to the colored noise case
  by computing the correlation in the frequency domain and dividing by
  the noise power spectral density function.} is $C_n=\int
x_n(t)\tilde{h}^*(t-\tau_n) dt$, denoting $x_n(t)$ the strain data at
detector $n$ and $\tilde{h}(t)$ the expected waveform defined above normalized
to unit ${\cal L}_2$ norm. The operator $\vec{P}=(P)_{mn}$ projects
onto the vector space generated by the antenna pattern vectors
$\vec{F}=(F)_{n}$ and $\vec{F}^*$.

The second derivatives of the likelihood ratio can be
calculated\footnote{As in \cite{Pai:2000zt}, we assume that the
  variation of the likelihood ratio is dominated by the effect of the
  time delays. We thus neglect the contribution from the phase shift
  due to the varying antenna pattern.} following \cite{Pai:2000zt}
(Sec. V A). The Hessian matrix reads
\begin{equation}
\label{eq:metric}
\partial^2_{\alpha\beta}L|_{\phi_0,\theta_0} = -\sum_{n,m=1}^D \Re\{p_{nm}\} g^{nm}_{\alpha\beta}.
\end{equation}
where the weights are $p_{mn}\equiv P_{mn}F_nF^*_m$.  The metric
components can be calculated explicitly
\begin{equation}
g^{nm}_{\alpha\beta}= 
\int\!\!\!\int {\cal A}^2(t){\cal A}^2(s) 
\partial_\alpha(\dot{\varphi}(t)\tau_n-\dot{\varphi}(s)\tau_m)
\partial_\beta(\dot{\varphi}(s)\tau_n-\dot{\varphi}(t)\tau_m)\:dt ds
\end{equation}
by invoking\footnote{We used the convention $\tau_1=0$.} slow
variations of the amplitude as to compared to the phase's $\dot{A}/A
\ll \dot{\varphi}$ and denoting ${\cal A}(t)$ the
signal amplitude after whitening (i.e., division of the spectrum by $\sqrt{S(f)}$
where $S(f)$ designates the PSD of
the instrumental noise which we assume identical at all detectors).

If the burst is monochromatic $\varphi(t)=2\pi f_0 t +\varphi_0$
with a constant amplitude $A(t)=1$
(which the case we consider in the simulations presented in the next
sections), this simplifies to
\begin{equation}
g^{nm}_{\alpha\beta}= {\cal A}_0^2 
[\partial_\alpha \cdot \partial_\beta] 
\left(2\pi f_0 (\tau_n - \tau_m)\right),
\end{equation}
where ${\cal A}_0\propto 1/S(f_0)$. It is interesting to note that the diagonal
terms $g^{nn}_{\alpha\beta}$ are zero so that the Hessian
matrix in Eq. (\ref{eq:metric}) vanishes when the projection operator 
is diagonal. This is true for instance when the number of detectors is $D=2$
as $\vec{P}=\vec{I}$ in that case (in words, this is equivalent to saying
that it is not possible to locate the GW source from two detectors only).

In the case of co-aligned (but not colocated) detectors (with
identical antenna patterns $F$), the above expression are further
simplified. We realize that this configuration is unrealistic as the
detectors are built on a spherical Earth. However, it provides us a
case where calculations and simulations are simpler. In this case,
$p_{mn}=|F|^2/D$
\begin{equation}
\partial^2_{\alpha\beta}L|_{\phi_0,\theta_0} = -{\cal A}_0^2 \frac{|F|^2 }{D} (2\pi f_0)^2 \sum_{nm}\partial_\alpha (\tau_n - \tau_m) \partial_\beta (\tau_n - \tau_m)
\end{equation}

With the above Taylor expansion, we can estimate the SNR loss due to a
mismatch between the pointing direction and the exact source
direction. Let $\mu$ denote the tolerable fraction of amplitude SNR
loss. The grid cell is thus defined by
\begin{equation}
\sum_{\alpha,\beta}\partial^2_{\alpha\beta}L|_{\phi_0,\theta_0} \Delta \alpha \Delta \beta
\leq 2 L(\phi_0,\theta_0) \left( 1-\mu^2 \right)
\label{e:ellipse_likelihood}
\end{equation}
with $L(\phi_0,\theta_0)={\cal A}^2 |F|^2$.

\begin{figure}
\begin{center}
  \includegraphics[width=\textwidth]{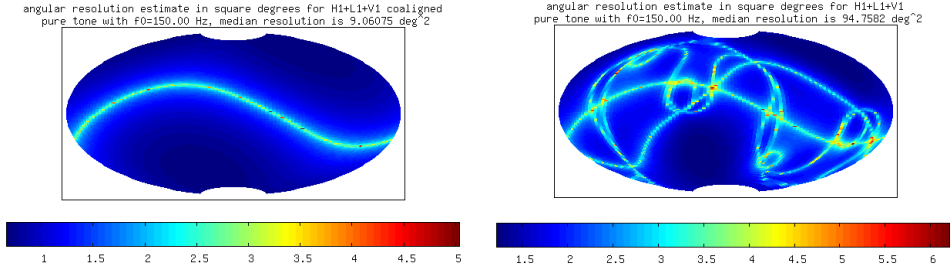}
\end{center}
\caption{\label{fig:resolution}\textbf{Estimate of the angular
    resolution given by the ($\log_{10}$) solid angle (in square
    degrees) subtended by the ellipse defined in
    Eq. (\ref{e:ellipse_likelihood})} for the 3 detector network A in
  Table \ref{t:number_samples} assuming coaligned detectors (left) and
  real orientation of the detectors (right). The signal is a pure tone
  of frequency $f_0=150$ Hz. In the first case, the angular resolution
  degrades when the coming wave has a small incidence w.r.t the
  detector plane (a similar effect appears with the classical
  triangulation method). The situation is clearly more complicated in
  the second case: the points where the resolution degrades form
  complicated patterns resulting from the interplay between the
  relative positions and orientations of the detectors. Note that the
  median resolution is about 9 square degrees when the detectors are
  assumed to be coaligned, while it is 10 times larger, about 94
  degrees when considering their true orientation.}
\end{figure}

This equation represents an ellipse centered at $(\phi_0,\theta_0)$.
The solid angle subtended by the ellipse expresses the \textit{angular
  resolution}\footnote{Note that the present use of the term ``angular
  resolution'' is somewhat non-standard. Here, we define the width of
  the peak relatively to its maximum. It is therefore independent of
  the amplitude of the gravitational wave. This quantity is equal to
  what is referred to as ``angular resolution'' up to a scaling factor
  depending on the signal-to-noise ratio and $\mu$.} achievable at
this particular sky position.  In Fig. \ref{fig:resolution}, we show
the (logarithm of the) angular resolution computed for a pure tone.
It is useless to sample the celestial sphere with a bin size smaller
than this ellipse if the goal is to get a fractional SNR loss of at
most $\mu$. Note that while this requirement (with typically
$\mu\approx 95\%$) is adequate for detecting a potential source, it is
probably not sufficient for estimating its sky position.

\subsection{Smallest grid extraction as a covering problem}

We want to cover the sky sphere with the smallest number of adjacent
sky resolution ellipses. Let $\Omega_e=\{s_k=(\phi_k,\theta_k)\}$ be the
sky grid obtained from Sec.~\ref{sec:td_sphangles} and let
$\mathcal{E}_k$ be the set of samples of $\Omega_e$ located inside the
sky resolution ellipse defined in Eq.~(\ref{e:ellipse_likelihood})
associated to $s_k$.

We define that two grid points $s_j$ and $s_k$ are \textit{adjacents} if $s_j \in
\mathcal{E}_k$ \textit{and} $s_k \in \mathcal{E}_j$. Note that the
symmetry of this criterion is important as $s_j \in \mathcal{E}_k$
does \textit{not} imply that $s_k \in \mathcal{E}_j$ since the sky
resolution ellipses change from points to points.

We search for the smallest subset of samples $\Omega_s$ such that all
samples in $\Omega_e$ are adjacent to at least one sample in
$\Omega_s$. This problem is similar to the template placement problem
encountered for the search of coalescing binaries
\cite{arnaud03,beauville05} but there are two major differences. First,
an initial sky-grid $\Omega_e$ is imposed here. Second, the topology of the
parameter space to cover is different: the periodicity of the sphere
makes the covering more difficult. In fact, there is no known optimal
solution to this problem in the general case. Only an approximated solution can
be constructed.

\subsection{Solving the covering problem by a greedy procedure}

The problem we have at hand can be formulated as an integer
optimization problem with linear constraints (see~\ref{a:scp},
Eq.~(\ref{eq:P})). In this form, it is a particular case of the
\emph{Set Covering Problem} (SCP)
\cite{christofides75,baker81,beasley87} which is known to be
NP-complete \cite{karp72}: no polynomial-time algorithm can solve
it. Many methods have been proposed (for a survey, see
\cite{caprara00}) and we selected the ones of \cite{beasley90} and
\cite{ceria98} that can handle large number of variables (typically,
the size of $\Omega_e$ is $\sim 10^4$ to $10^5$) and constraints
(equal to the square of the initial grid size). We describe the
algorithm in details in \ref{a:scp}.  It consists of an efficient
greedy procedure based on a dual Lagrangian relaxation of the
problem. The greedy procedure iteratively builds a good solution: at
each iteration, the variable $x_k$ that maximizes a specific cost
function is added to the solution and all rows $j$ such that $x_j$ and
$x_k$ are adjacents are removed from the problem, producing a new
reduced problem that can be treated in a similar way.  In \ref{a:scp},
we highlight the modifications we made to adapt to the present
problem. In particular, we use a cost function based on the
information obtained from the sole dual Lagrangian relaxation of the
linear problem.

\subsection{Application}

In this section, we illustrate the method we propose by applying it to
two different networks. We consider fictitious networks composed of
detectors with parallel orientations. We can check our results against
estimates of what should be the grid size in this case. We also
consider realistic networks using the actual position and orientation
of the main detectors currently in operation.

From Eq.~(\ref{eq:metric}), we get the sky resolution ellipses for
every samples assuming that the GW has frequency $f_0=150$ Hz.  This frequency 
approximately corresponds to the best sensitivity for LIGO and Virgo detectors. 
We build the corresponding
adjacency matrix which feeds the greedy algorithm based on the dual
Lagrangian relaxation.

Considering that there are approximately 41,253 square degrees in the
whole sphere. From the median angular resolution obtained in
Fig. \ref{fig:resolution}, we expect the full sky coverage with a grid of
4500 vertices in the case with coaligned detectors and about 450 vertices 
considering the true orientation of the detectors.

\paragraph{Detectors with parallel orientations}

\begin{figure}
\begin{center}
  \includegraphics[width=.45\textwidth]{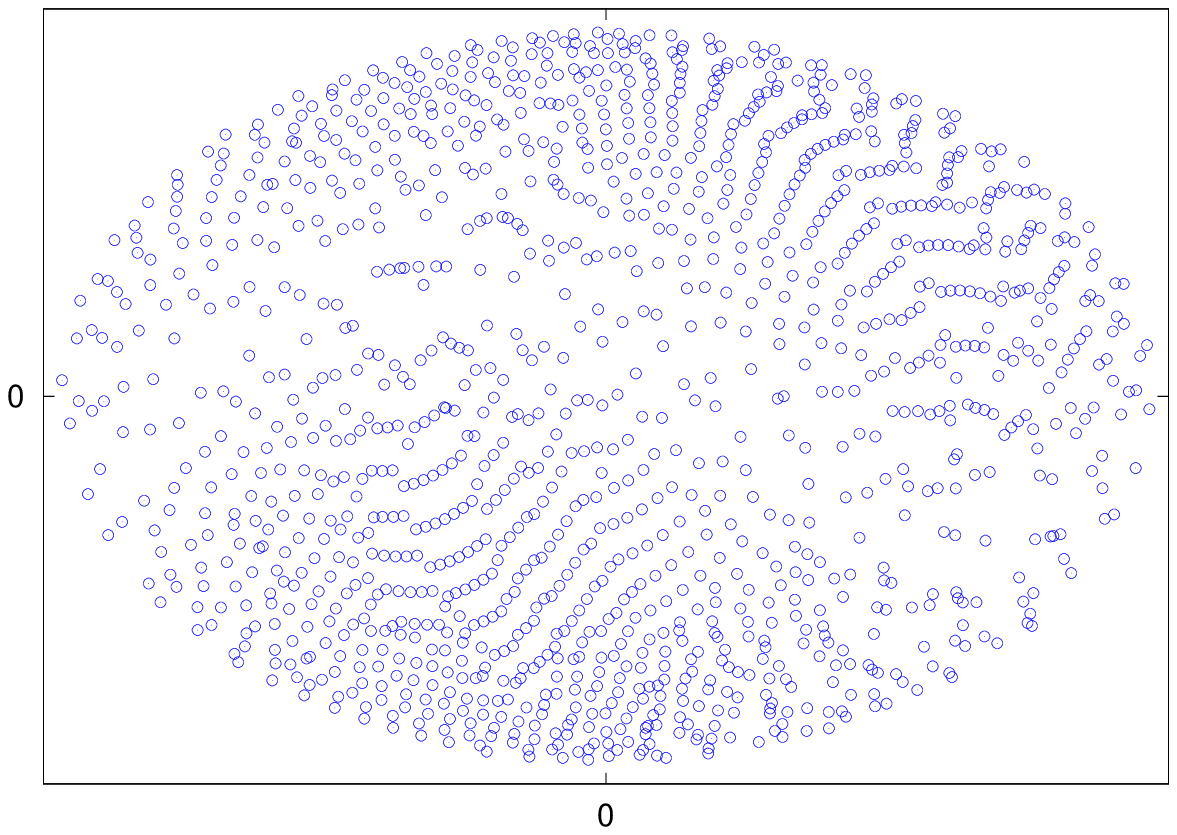}
  \includegraphics[width=.45\textwidth]{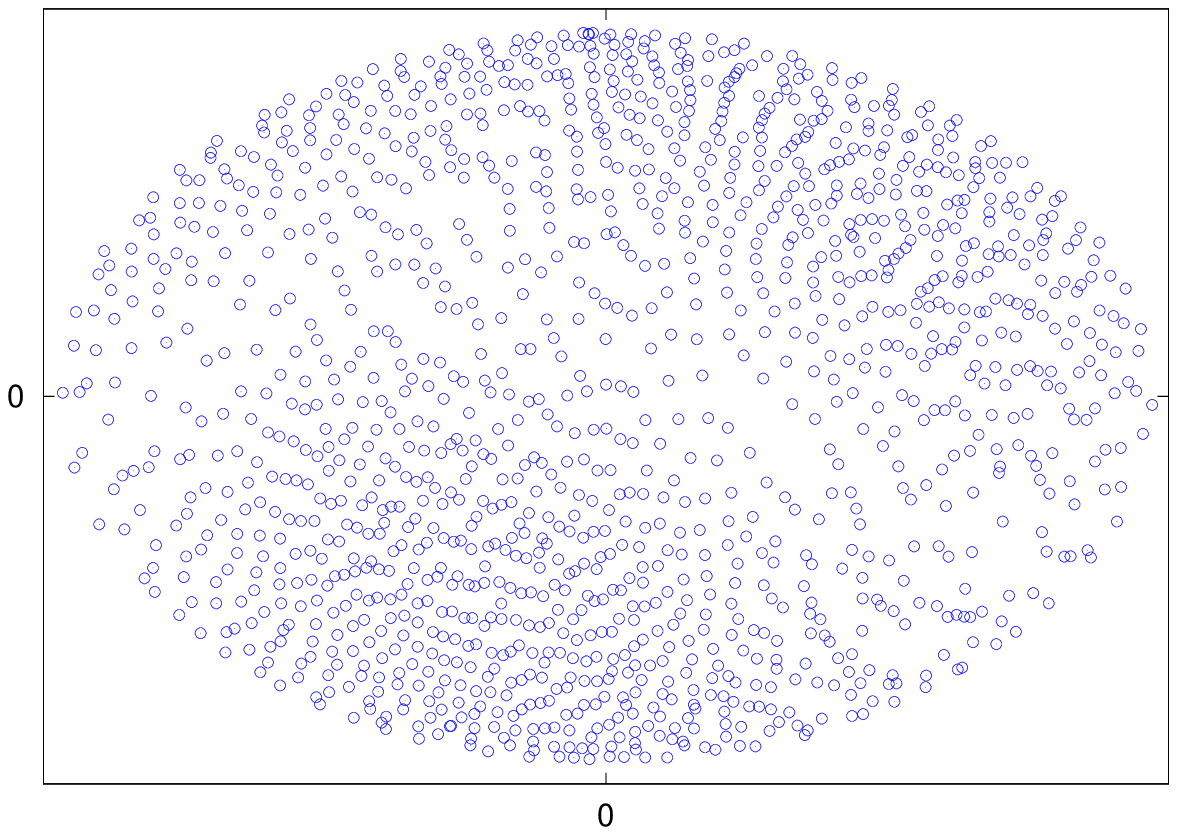}\\
  \includegraphics[width=.45\textwidth]{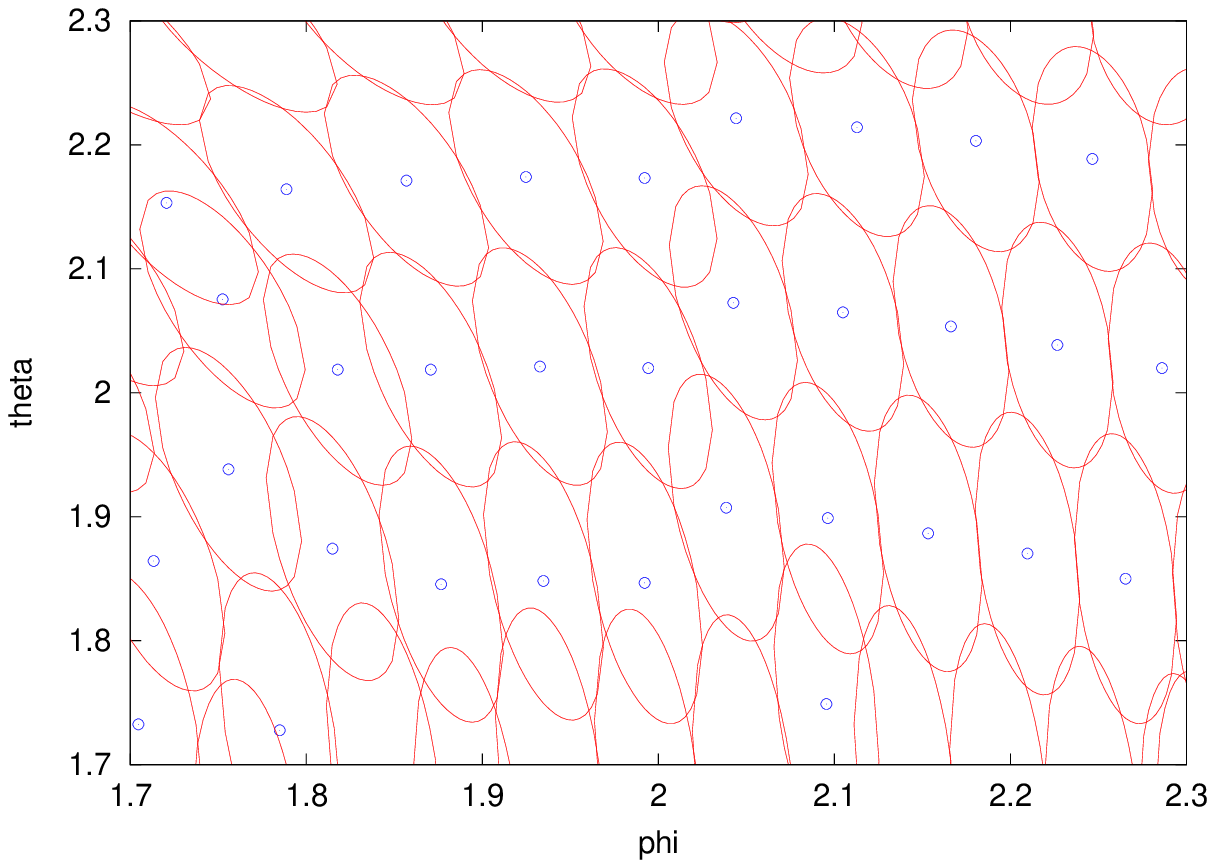}
  \includegraphics[width=.45\textwidth]{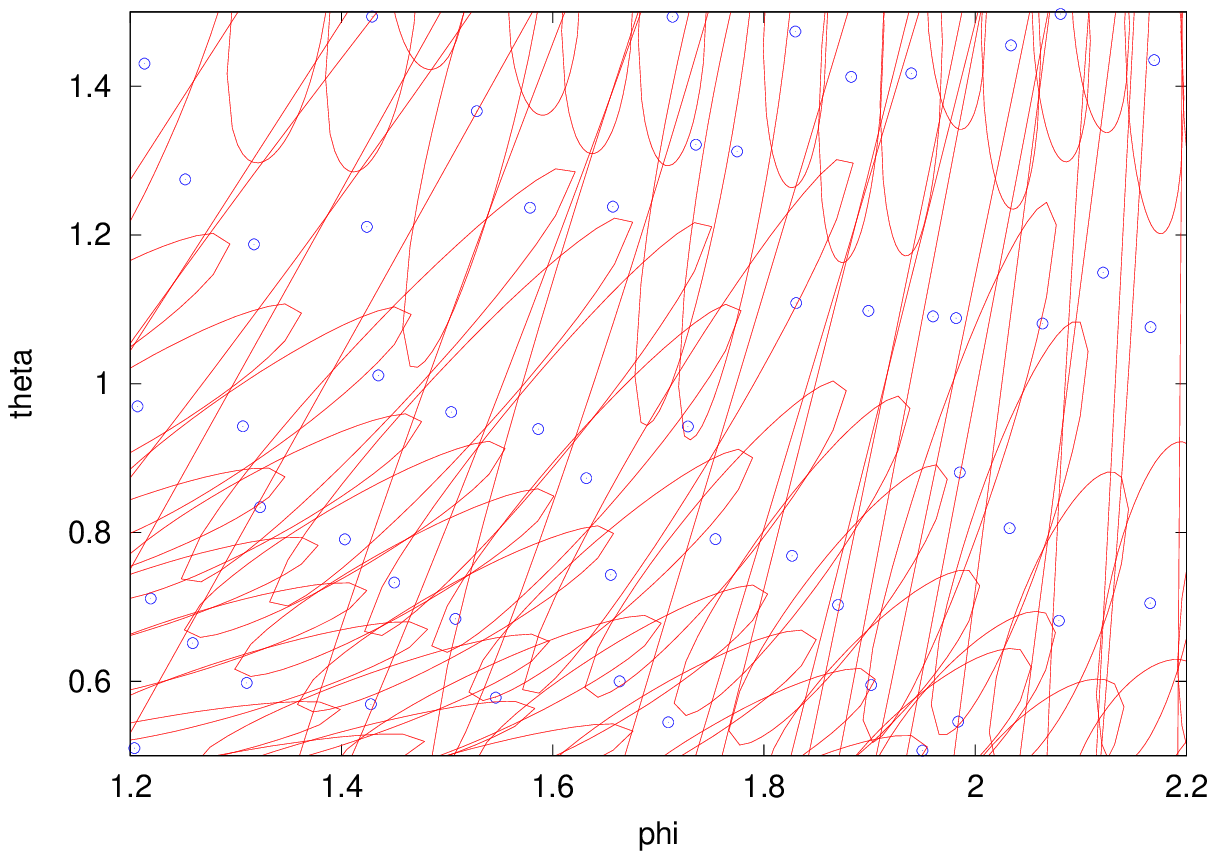}
\end{center}
\caption{\textbf{Minimal sky grids for the
    3 detector network A (top left, see Table \ref{t:number_samples}
    for the description of the networks) and 4 detector network B (top
    right) assuming parallel orientations (sampling frequency
    $f_s=4096$ Hz) and signal frequency $f_0=150$ Hz.}  The number of
  vertices in the grid is $1284$ in the first case and $1354$ in the
  second. Bottom row, we present two blow-up for the 3 detector case:
  (left) where the sky resolution ellipses have constant size and
  remain aligned and the other; (right) where the ellipses are
  strechted and vary rapidly from samples to samples.}
\label{fig:full_sky_best_samples_fs_4096_fc_150} 
\end{figure}

We first examine the three-detector network A in Table
\ref{t:number_samples} (Virgo, Ligo H and Ligo L, assuming identical
noise PSDs and identical orientations) and four detector B (adding
GEO). The sampling frequency is set to $f_s=4096$ Hz.  The size of
initial sky grid produced by the procedure in
Sec.~\ref{s:td_to_sky_grids} is $|\Omega_e|=28153$. The set covering
problem can be qualified as large as the constraint matrix in
Eq. (\ref{eq:P}) is of size $|\Omega_e| \times |\Omega_e|$. The
application of the greedy algorithm based on the dual lagrangian
relaxation results in the sky grid presented in
Fig.~\ref{fig:full_sky_best_samples_fs_4096_fc_150} which contains
$|\Omega_s|=1284$ vertices, i.e. $\simeq 4.6\%$ of the initial number
of samples. This result is consistent with that of \cite{Pai:2000zt}.
Rescaling the results obtained for case III of Table (II) to the
special case discussed here, we get a sky grid of size $n_s^\Omega
\sim 1780$ which is comparable to $|\Omega_s|$ mentioned above.

This figure also displays a zoom of the resulting grid along with the
corresponding ellipses in two small areas. In the area where the
ellipse shape and orientation remain roughly constant, we see that
the coverage is regularly performed by adjacent ellipses. In area
where the ellipses are stretched and have orientation that changes
rapidly (this is the case close to the line $\phi=0$ in the detector
plane), some degree of overlap is required to have the complete
coverage. This case illustrates the importance of the symmetric
adjacency proposed and used here.

The minimal sky grid has been computed with various detector networks
and various sampling frequencies. The results are tabulated in
Table~\ref{t:minimum_set_sizes}. As expected the size of the minimum
grid remains more or less constant with the sampling frequency, as it
is prescribed by the angular resolution achievable by the detector
network.

\begin{table}
\centering
\begin{tabular}{|c|c|c|c|c|}
\hline
\multicolumn{2}{|c|}{$f_s$ (Hz)} & 1024 & 2048 & 4096 \\
\hline
\hline
\multirow{3}{*}{A: 3-det} & exh. set ($|\Omega_e|$)    & 1,868 & 7,184 & 28,153 \\
 \cline{2-5}
                               & min. set ($|\Omega_s|$)    & 1,774 & 1,127 & 1,284 \\
 \cline{2-5}
                               &  $|\Omega_s|/|\Omega_e|$ ($\%$) & 95.0 & 15.7 & 4.6 \\
\hline
\hline
\multirow{3}{*}{B: 4-det} & exh. set ($|\Omega_e|$)     & 3,983 & 16,650 & 66,454 \\
 \cline{2-5}
                               & min. set ($|\Omega_s|$)    & 1,454 & 1,180  & 1,354 \\
 \cline{2-5}
                               & $|\Omega_s|/|\Omega_e|$ ($\%$) & 36.5 & 7.1   & 2.0 \\
\hline
\hline
\multirow{3}{*}{C: 5-det} & exh. set ($|\Omega_e|$)   & 22,985 & 90,740 & 351,171 \\
 \cline{2-5}
                               & min. set ($|\Omega_s|$)   & 2,695 & 2,979 & 3,368 \\
 \cline{2-5}
                               & $|\Omega_s|/|\Omega_e|$ ($\%$) & 11.7 & 3.3 & 0.96 \\
\hline
\end{tabular}

\caption{\textbf{Comparison between the sizes of initial sky grid
    $\Omega_e$ and minimum sky grid $\Omega_s$} assuming a parallel 
  orientation for the detectors. We get a typical size of few thousands of vertices 
  which is consistent with the rough estimate drawn from the median resolution in 
  Fig. \ref{fig:resolution}.}
\label{t:minimum_set_sizes}
\end{table}

To check that good detection performance can still be achieved with
the minimum grid, we also compared the performance in terms of false
alarm and detection probabilities of both the initial grid and the
minimum grid. False alarm and detection rates were evaluated by Monte
Carlo method using noise only (using simulated white Gaussian noise)
and signal+noise trials. A monochromatic signal with frequency
$f_0=150$ Hz was injected in the 3-detector network A using a sampling
frequency $f_s=4096$ Hz. The position of the source was uniformly
drawn over the sky sphere. The other signal parameters such as
polarization or amplitude were also randomly chosen. 

We used the (-log) likelihood ratio introduced in
Sec. \ref{s:angular_resol} as the detection statistics following the
implementation given in \cite{pai08}. This corresponds to applying a
matched filter filtering procedure which results in a likelihood sky
map. The maximum of this likelihood map was compared to a threshold to
decide for the detection of the GW.

\begin{figure}
\begin{center}
  \includegraphics[width=.6\textwidth]{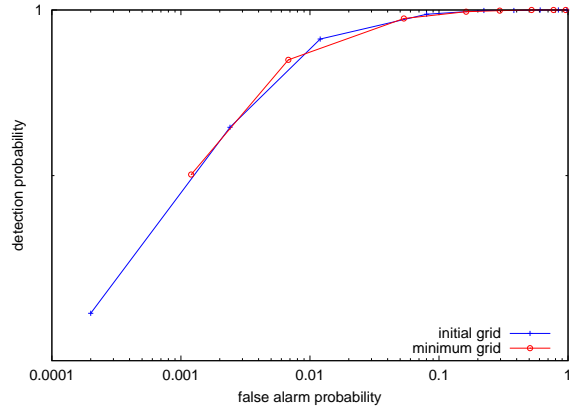}
\end{center}
\caption{\textbf{Detection performance in terms of false alarm and detection probabilities} for the 3-detector network A (sampling frequency $f_s=4096$). }
\label{fig:log_roc_exh_vs_min_set} 
\end{figure}

These simulations were repeated for different thresholds. For
threshold values, $10^4$ Monte Carlo simulations were run. The result
obtained is presented in Fig.~\ref{fig:log_roc_exh_vs_min_set}. We
observe that the performances obtained with the minimal sky grid are
comparable to that of the (``oversampled'') initial sky grid.

\paragraph{Detectors with their true orientations}

\begin{figure}
\begin{center}
  \includegraphics[width=.5\textwidth]{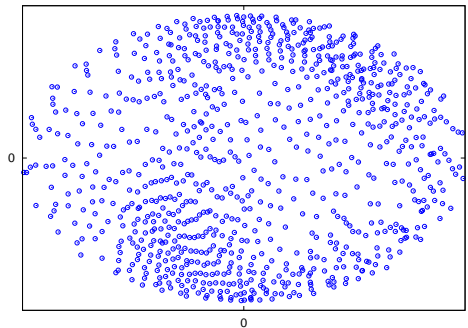}
\end{center}
\caption{\textbf{Minimal sky grid ($887$ vertices)
  for the 4 detector network B (see Table \ref{t:number_samples})
  Virgo, Ligo H, Ligo L and GEO with their actual orientation
  (sampling frequency $f_s=4096$ Hz and signal frequency $f_0=150$
  Hz.)}}
\label{fig:full_sky_best_samples_4_det_real_metric_fs_4096_fc_150}
\end{figure}

Now, we consider the detectors with their true orientation. We show in
Fig.~\ref{fig:full_sky_best_samples_4_det_real_metric_fs_4096_fc_150}
the result we get for the four-detector network B in Table
\ref{t:number_samples}. The size of the minimum grids are gathered in
Table~\ref{t:minimum_set_sizes_real_network} for various
configurations.

The resulting grids contain less samples than in the previous case
(parallel detectors) due to the additional loss of accuracy which
arises when the detectors don't share the same orientation.

\begin{table}
\centering
\begin{tabular}{|c|c|c|c|c|}
\hline
\multicolumn{2}{|c|}{$f_s$ (Hz)} & 1024 & 2048 & 4096 \\
\hline
\hline
\multirow{3}{*}{A: 3-det} & exh. set ($|\Omega_e|$)    & 1,868 & 7,184 & 28,153 \\
 \cline{2-5}
                               & min. set ($|\Omega_s|$)    & 429 & 391 & 440 \\
 \cline{2-5}
                               &  $|\Omega_s|/|\Omega_e|$ (in $\%$) & 23.0 & 5.4 & 1.6 \\
\hline
\hline
\multirow{3}{*}{B: 4-det} & exh. set ($|\Omega_e|$)     & 3,983 & 16,650 & 66,454 \\
 \cline{2-5}
                               & min. set ($|\Omega_s|$)    & 937 &  817 &  887\\
 \cline{2-5}
                               & $|\Omega_s|/|\Omega_e|$ ($\%$) & 23.5 & 4.9 & 1.3 \\
\hline
\hline
\multirow{3}{*}{C: 5-det} & exh. set ($|\Omega_e|$)   & 22,985 & 90,740 & 351,171 \\
 \cline{2-5}
                               & min. set ($|\Omega_s|$)   & 1,587 & 1,723 & 1,898 \\
 \cline{2-5}
                               & $|\Omega_s|/|\Omega_e|$ ($\%$) & 6.9 & 1.9 & 0.54 \\
\hline
\end{tabular}
\caption{\textbf{Comparison between the sizes of initial sky grid
    $\Omega_e$ and minimum sky grid $\Omega_s$} considering the actual 
  orientation of the detectors. We get a typical grid size of few hundrends of vertices which is consistent with the estimate drawn from the median resolution in Fig. \ref{fig:resolution}.}
\label{t:minimum_set_sizes_real_network}
\end{table}

\section{Conclusions}
\label{s:conclusion}

Most coherent detection scheme includes an algorithmic loop over
source directions taken from a predefined sky grid. This part of the
detection algorithm takes a significant amount of computing
ressources. One of the available degrees of freedom to address this
issue is the way the sky sphere is sampled.  We investigated the
possibility to build an optimally ``sparse'' sky grid which performs
the complete sky coverage with the smallest number of vertices. It is
clear that the optimal sky grid should adapt the pointing accuracy.

Contrarily to electromagnetic antenna arrays (e.g., used for RADAR
applications), gravitational wave detector networks are not arrays of
regularly spaced and oriented sensors. The pointing accuracy we get
with such network results from the complicated interplay between the
relative positions and orientations of the detectors. Because of the
network heterogeneity, the pointing accuracy varies largely over the
whole sky. In this paper, we build sky grids from an estimate of the
``local'' angular resolution.  The vertex density is reduced where the
resolution is coarse and vice-versa.

The method goes into two steps. We first produce a grid from
pre-selected sky directions where the travel time between detectors
and the time reference is (close to) an integer number of time
samples. Those points are convenient because the time-shifts necessary
to align in time the signals (i.e., compensate the travel time between
the detector and the time reference) received by each detector are
trivially performed. This operation amounts to sampling the manifold
described by the travel times. This sampling can be determined
analytically in the case of three (i.e., the manifold is an ellipse
interior) and four detectors (i.e., the manifold is the surface of an
ellipsoid). For larger networks, we suggest to apply Monte Carlo
procedures. Once done, a key point is to map from travel times to sky
directions. To perform this mapping, we propose a method that is
robust to the round-off error due to the truncation of the travel
times.

In the second step, we extract from the first grid the smallest
sub-set of vertices that ensures complete sky coverage for a given
loss in signal-to-noise ratio.  We build this sub-set by casting the
grid size minimization into a \textit{set covering problem}.  The
procedure we propose to solve the corresponding linear program with
linear and integer constraints relies on a greedy algorithm and a dual
Lagrangian relaxation. The resulting grid show considerable reduction
in size. We have checked that the results we get with this grid are
comparable to what we get when using more resolved sky grids (i.e., at
a given false alarm rate, the detection probability are comparable
when performing an all-sky blind search). Our investigations have been
limited to source detection. It is likely that the sky grid has to be
refined if the goal is the accurate determination of the source
position.

The overall procedure to build the grid can be quite computationally
heavy. However this procedure needs to be done only once beforehand.

\section*{Acknowlegments}
This research was supported by the Virgo-EGO Scientific Forum.

\appendix
\section{Solving the set covering problem}
\label{a:scp}
We propose to formulate the problem of building the minimum sample set as an integer optimization problem with linear constraints in the following way:
\begin{equation}
\begin{array}{rl}
\min & \displaystyle{\vec{u}^T\vec{x}=\sum_{k=1}^{|\Omega_e|} x_k} \\
\text{s.t.} & \vec{Ax}\geq\vec{u}, \\
& x_k\in\{0,1\},~ k=1, \dots, |\Omega_e|.
\end{array}
%\tag{P}
\label{eq:P}
\end{equation}
Here the vector $\vec{x}$ of size $|\Omega_e|$ is the output vector that should define the minimal set: its entry $x_k$ takes value $1$ if the corresponding sample $s_k$ is included in $\Omega_s$, and $0$ otherwise. Hence, the non-zero entries of $\vec{x}$ define the subset $\Omega_s$, and the objective function of this optimization problem is the number of samples in the final set (the number of non-zero values of vector $\vec{x}$, to be minimized). The vector $\vec{u}$ is a column vector of size $|\Omega_e|$ such that $\vec{u}=[1,\dots,1]^T$. Finally the matrix $\vec{A}$ of size $(|\Omega_e| \times |\Omega_e|)$ represents the neighbouring relationships: it is such that $a_{ij}=1$ if $s_i$ and $s_j$ are neighboors and $a_{ij}=0$ otherwise. Each row $j$ of matrix $\vec{A}$ represent a specific constraint on sample $s_j$ in the oversampled set: it is here to insure that  $s_j$ is be adjacent to at least one non zero entry of $\vec{x}$, and therefore covered by at least one resolution cell.

As the matrix of linear constraints $\vec{A}$ may contain several tens or hundreds of thousands of columns and the same amount of rows (see the number of samples from Table~\ref{t:number_samples}, the problem is very difficult and cannot be solved by optimal methods. We therefore propose to use here a method inspired by \cite{beasley90} and \cite{ceria98}. It follows an efficient greedy procedure based on a dual Lagrangian relaxation of the problem. The greedy procedure iteratively builds a good solution in the same way as in \cite{ceria98}: at each iteration, the variable $x_k$ that maximizes a specific cost function is added to the solution and all rows $j$ such that $x_j \in \mathcal{E}_k$ are removed from the problem (because the corresponding constraints are fulfilled by the solution), producing a reduced problem that can be treated in a similar way. Here the cost function is computed thanks to the information obtained from the dual Lagrangian relaxation of the reduced problem. Contrary to the method proposed in \cite{ceria98}, we do not use the primal relaxation of (\ref{eq:P}), which reduces the computational cost of the algorithm as well as the dependency of the method over tunable parameters, and leads to smaller running time and slightly better solutions.

The dual problem of  (\ref{eq:P}) is defined as:
\begin{equation}
\begin{array}{rl}
\max & \vec{u}_m^T \vec{y}\\
\text{s.t.} & \vec{A}^T \vec{y} \leq \vec{u}_n, \\
& y_k \geq 0,~ k=1, \dots, |\Omega_e|.
\end{array}
%\tag{D}
\label{eq:D}
\end{equation}

The Lagrangian relaxation of the dual problem (\ref{eq:D}) is given here by:
\begin{equation}
\begin{array}{rl}
\displaystyle{\min_{\pmb{\mu}}} &  \displaystyle{\max_{\vec{y}}L(\vec{y},\pmb{\mu})} \\
\text{s.t.} & \mu_i\geq 0,~ i=1, \dots, m, \\
& 0 \leq y_k\leq 1,~ k=1, \dots, n,
\end{array}
%\tag{LD}
\label{eq:LD}
\end{equation}
where $\pmb{\mu}$ is the vector of Lagrange multipliers and $L(\vec{y},\pmb{\mu}) = \vec{u}_m^T\vec{y} - \mu^T(\vec{A}^T\vec{y}-\vec{u}_n)$ is the Lagrange function of the dual problem. Here the maximization of $L(\vec{y},\pmb{\mu})$ over $y_k\in\{0,1\},~ k=1, \dots, n$ is straightforward: indeed it is obtained by setting $y_k=1$ if the $k^{\text{th}}$ column of the row vector $\vec{u}_m^T - \mu^T\vec{A}^T$ is positive, and $y_k=0$ otherwise. Besides the function $\max_{\vec{y}}L(\vec{y},\pmb{\mu})$ is convex in $\pmb{\mu}$, and therefore it can be minimized efficiently by means of a subgradient algorithm. Two main problems arise:
\begin{itemize}
\item the optimal values of (\ref{eq:D}) and of its Lagrangian relaxation (\ref{eq:LD}) may not be equal (duality gap);
\item the vector $\vec{y}^*(\pmb{\mu})$ minimizing $L(\vec{y},\pmb{\mu})$ is generally not feasible, i.e. it doesn't verify the linear constraints $\vec{A}^T\vec{y}^*(\pmb{\mu})\leq\vec{u}_n$.
\end{itemize}

However, despite these problems, and contrary to the multipliers obtained from the primal Lagrangian relaxation \cite{beasley90}, the dual Lagrange multipliers $\pmb{\mu}$ carry very useful information about problem (\ref{eq:P}). In particular $\pmb{\mu}$ is a near-optimal and near-feasible solution of the linear relaxation of (\ref{eq:P}). It can therefore be used to decide which variable should be added to the final solution. Here the variable chosen is simply the one corresponding to the largest entry of $\pmb{\mu}$. It should be also noted that all samples $s_k$ such that $\mathcal{E}_k=\{ s_k \}$ must be in the final solution set since these points have no adjacent point. 

The final greedy algorithm can be summarized as follows:
\begin{enumerate}
\item Find the oversampled set $\Omega_e$;
\item Build the adjacency matrix $A$;
\item Set the solution set $\Omega_s=\emptyset$;
\item Find all points $s_k$ such that $\mathcal{E}_k=\{ s_k \}$, add them to $\Omega_s$, remove them from the set $\Omega_e$ and remove the corresponding rows and columns from the matrix $\vec{A}$;
\item While $\Omega_e \not= \emptyset$:
\begin{itemize}
\item Solve the dual Lagrangian relaxation (\ref{eq:LD}) of problem (\ref{eq:P});
\item Find the largest Lagrange multiplier $\mu_k$, add the corresponding sample $s_k$ to $\Omega_s$, remove the corresponding column from matrix $\vec{A}$ as well as all rows $j$ such that $s_j$ and $s_k$ are neighboors, and reduce the set $\Omega_e$ similarly.  
\end{itemize}
\end{enumerate}

It is also sometimes possible to improve the quality of the solution
by introducing a randomized criterion for the variable selection. In
\cite{ceria98}, it was proposed to run several time the algorithm. For
the first run, the variable selected is, as described above, the one
corresponding to the largest entry of $\pmb{\mu}$. However for the
other runs, the variable selected may be the one corresponding to the
second largest entry of $\pmb{\mu}$ with probability $0.5$. We adopted
this strategy since the choice of the minimum sample set has to be
done only once in our case and therefore the additional computational
cost induced by running several time the same algorithm is not
prohibitive.

\section*{References}
\bibliographystyle{unsrt}
\bibliography{paper}

\end{document}